\documentclass[twocolumn,prb,aps,floatfix,longbibliography,superscriptaddress]{revtex4-2}

\usepackage{color}
\usepackage{graphicx}
\usepackage{siunitx}
\usepackage[color=green!60,textsize=small]{todonotes}
\usepackage{physics}
\usepackage{amsthm}
\usepackage{amsmath}
\usepackage{amssymb}
\usepackage{enumerate}
\usepackage{placeins}
\usepackage{mathtools}
\usepackage{booktabs}
\usepackage{dsfont}
\usepackage{yfonts}
\usepackage{calligra}
\usepackage{hyperref}
\usepackage{svrsymbols} 
\usepackage{tikz}
\usepackage{enumitem}
\usepackage[normalem]{ulem}

\DeclareMathAlphabet{\mathcalligra}{T1}{calligra}{m}{n}
\DeclareFontShape{T1}{calligra}{m}{n}{<->s*[2.2]callig15}{}

\renewcommand{\vec}[1]{\boldsymbol{#1}}
\newcommand{\Eqref}[1]{Eq.~\eqref{#1}}

\usepackage{array}
\usepackage{hhline}
\usepackage{latexsym}
\usepackage{bm}
\usepackage{bbm}
\usepackage{cancel}

\newcommand{\Vext}{\mathcal{V}_{\rm He-\graphene}}
\newcommand{\Vhehe}{\mathcal{V}_{\rm He-He}}

\AtBeginDocument{%
\heavyrulewidth=.08em
\lightrulewidth=.05em
\cmidrulewidth=.03em
\belowrulesep=.65ex
\belowbottomsep=0pt
\aboverulesep=.4ex
\abovetopsep=0pt
\cmidrulesep=\doublerulesep
\cmidrulekern=.5em
\defaultaddspace=.5em
}

\begin{document}

\title{A Perspective on Collective Properties of Atoms  on 2D Materials}

\author{Adrian Del Maestro}
\affiliation{Department of Physics and Astronomy, University of Tennessee, Knoxville, TN 37996, USA}
\affiliation{Min H.~Kao Department of Electrical Engineering and Computer Science, University of Tennessee, Knoxville, TN 37996, USA}
\affiliation{Department of Physics, University of Vermont, Burlington, VT 05405, USA}

\author{Carlos Wexler} 
\affiliation{Department of Physics and Astronomy, University of Missouri, 
Columbia, MO 65211, USA}

\author{Juan M.~Vanegas}
\affiliation{Department of Physics, University of Vermont, Burlington, VT 05405, USA}

\author{Taras Lakoba} 
\affiliation{Department of Mathematics \& Statistics, University of  Vermont, Burlington, VT 05405, USA}

\author{Valeri N.~Kotov}
\affiliation{Department of Physics, University of  Vermont, Burlington, VT 05405, USA}

\begin{abstract}

Atoms deposited on two-dimensional (2D) electronic materials, such as graphene, can  exhibit unconventional many-body correlations, not accessible in other settings.  All of these are driven by van der Waals forces: between the atoms themselves and atom-material interactions.  For example $^4$He atoms on 2D materials can potentially form a variety of exotic quantum states of matter, such as two-dimensional supersolids and superfluids, in addition to solid phases. For the ``most quantum" case of a single helium layer we discuss, from a theoretical perspective, how the effective low-energy (Bose-Hubbard) description  can take advantage of the extreme sensitivity of this unique system  to the interplay between the atomic (helium) and solid-state (graphene) components. Due to the extraordinary variety and tunability of 2D electronic materials, we envisage that  a wide range of  correlated atomic phases can be  realized under favorable conditions. We also outline exciting possibilities in the opposite extreme of many atomic layers forming a liquid on top of graphene --  in this case  a so-called ``spinodal de-wetting"  pattern can form at the liquid--vapor interface which reflects the presence and electronic properties  of graphene underneath. Such patterns could  be manipulated by choosing different atoms and materials, with potential technological applications.

\end{abstract}

\maketitle






\section{van der Waals Interactions as a source of many-body collective behavior}
 The strength and range of two-particle interactions, combined with particle statistics,  are essential ingredients in determining many-body collective behavior. This includes the possibility of complex collective states of matter, such as superconducting/superfluid phases and correlated Mott insulators.  Under the right conditions, which  usually depend on  changes in parameters such as electron density, lattice structure, application of pressure, etc., quantum phase transitions can take place between correlated  states with different symmetries at zero temperature \cite{sachdev2011}.  Since it is extremely difficult to exactly take into account strong interaction effects, a fruitful approach is to develop low-energy effective descriptions which correctly predict the nature of different phases. 

Interactions between neutral atoms and materials, and between pairs of atoms, are of van der Waals (VDW) nature \cite{Israelachvili}.
They control a wide variety of physical phenomena ranging from extreme quantum behavior of atomic gases near material surfaces \cite{Nichols:2016hd,Yu:2021tw}, to
collective properties of thin liquid films  forming on material substrates \cite{Sengupta}.  In this perspective, we describe these two regimes of many-body behavior, and various routes towards manipulating it,  for the benefit of improving our theoretical understanding of correlated phenomena, and potentially exploiting it for novel future technologies. 

An important and unique  feature of atoms on 2D materials, such as graphene,  is the opportunity of exploiting the great tunability of their properties.  We thus envisage this system to be a novel platform for studies of atomic many-body states and quantum phase transitions directly in a solid-state setting.  Quantum effects are most important for the first layer of atoms, while for many atomic layers forming a liquid,  the VDW forces control wetting phenomena and surface instabilities, such as spinodal de-wetting.  

When we consider atoms on top of graphene for example, it is evident that the extreme  sensitivity of VDW interactions to the underlying lattice and electronic properties  is unique to solid-state environments where the atom-atom  two-body potential range and the lattice spacing are comparable.  Consequently, one  can follow a variety of directions  to manipulate 2D material properties, which will in turn significantly alter atom--2D material  potentials. For example, graphene and other 2D materials, such as members of the dichalcogenides family (MoSe$_2$, MoS$_2$,  WSe$_2$, WS$_2$), can withstand mechanical strain of at least twenty percent \cite{Lee385,Cooper-Hone}, leading to substantial changes in their electronic properties \cite{Antonio,Naumis_2017,maria,Nichols:2016hd}.  Changes in the electronic  environment lead to changes in polarization properties of the 2D layers, and therefore ultimately manifest themselves in  different van der Waals potentials that atoms experience close to the 2D sheets.  The  dichalcogenides are already quite different from graphene, as they exhibit an electronic gap and thus weaker polarization.  2D  materials can also be arranged geometrically to achieve desired properties \cite{vdwhetero,vd1}.  The possibilities for novel electronic states are almost boundless: for example, twisted  graphene bilayers at ``magic" angles can become strong Mott insulators and even unconventional superconductors  \cite{Cao1-2018,Cao2-2018}. Tendencies towards strong electronic localization in twisted graphene layers, or for  graphene on  structurally similar substrates \cite{Uchoa,Seo}, would inevitably result in different atom--material interactions.  We propose a research direction based on the realization that  changes in the local lattice and electronic structure due to external factors, or working with different members of the 2D material family, would result in substantially different VDW forces and by extension a greater variety of many-body behavior.


\section{Atoms near Graphene}

A system of $N$ indistinguishable atoms located at $\qty{\vec{r}_1,\dots,\vec{r}_N}$ proximate to a pristine graphene membrane can be described by the microscopic Hamiltonian
\begin{equation}
    H = -\frac{\hbar^2}{2m} \sum_{i=1}^N \nabla_i^2 + \sum_{i=1}^N \mathcal{V}_{\rm He-\graphene}(\vec{r}_i) + 
    \sum_{i <j} \mathcal{V}_{\rm He - He}(\vec{r_i}-\vec{r_j}) 
\label{eq:Ham}
\end{equation}
where the interaction between the atoms and graphene $(\Vext)$ as well as that between atoms $(\Vhehe)$ are of van der Waals origin.  In this perspective, we focus on $^4$He adsorbates.  They represent an ideal candidate, as they are light, highly symmetric (thus having rather weak VDW interactions) and exhibit quantum behavior, including superfluidity in the bulk.  While an empirical form of $\Vhehe$ has been determined to high accuracy by fitting to known experimental results \cite{Przybytek:2010ol,Cencek:2012iz,heprops}, the adsorption potential $\Vext$ has greater uncertainty \cite{Bruch:2007bk,Nichols:2016hd}, with a commonly used form involving the sum of individual spherically isotropic $6-12$ Lennard--Jones interactions between $^4$He and C. The long-range part (i.e. the attractive potential tail) of   $\Vext(z) \sim - z^{-4}$, in the intermediate distance range, has been thoroughly investigated  \cite{Churkin,Chaichian,Nichols:2016hd}. It is also quite  sensitive to external factors such as strain and chemical potential (electron density).

%
\begin{figure}[t]
\begin{center}
    \includegraphics[width=\columnwidth]{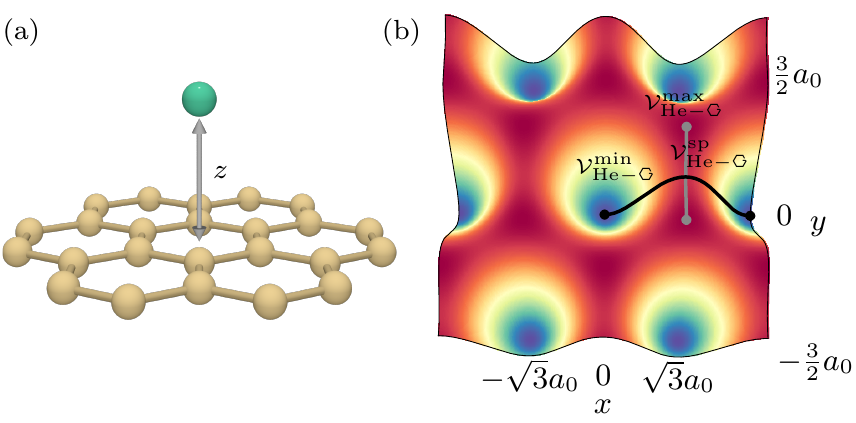}
\end{center}
\caption{(a) A helium atom located a distance $z$ above a graphene hexagon center which form a triangular lattice. (b) The helium--graphene adsorption potential $\Vext$ at its minimum value $z \approx \SI{3}{\angstrom}$ above the membrane.  The potential can be approximated using only 3 numbers: the distance between the minimum and maximum $(\Vext^{\rm max} - \Vext^{\rm min}) \sim \SI{20}{\kelvin}$ and that between the saddle point and maxima $(\Vext^{\rm max} - \Vext^{\rm sp}) \sim \SI{20}{\kelvin}$). It's overall minimum $\Vext^{\rm min} \approx \SI{-150}{\kelvin}$ controls the binding energy.}
\label{fig:SingleAtom}
\end{figure}
%

\begin{figure}[t]
\begin{center}
    \includegraphics[width=\columnwidth]{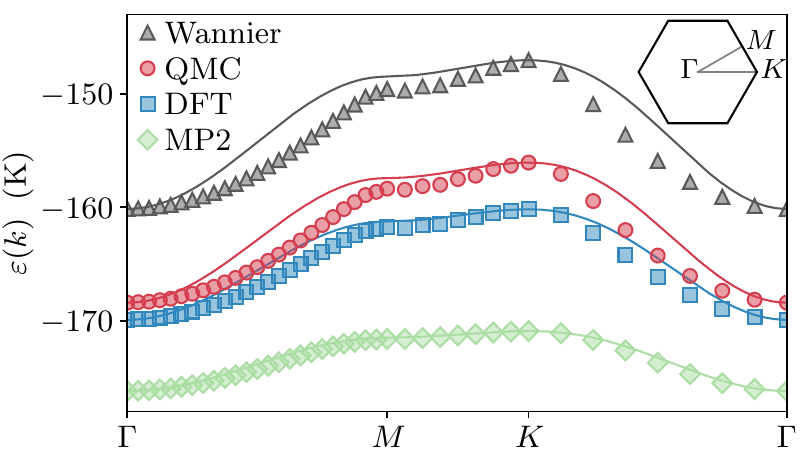}
\end{center}
\caption{The single-particle lowest energy bands computed from an effective He-graphene adsorption potential determined from the empirical potential in \Eqref{eq:Ham} (Wannier), or by computing adsorption energies in the plane via quantum Monte Carlo (QMC), density functional theory (DFT) and M\o{}ller--Plesset perturbation theory (MP2) \cite{Yu:2021tw}.  Lines show the tight-binding prediction in \Eqref{eq:tbdispersion} without any fit parameters using $t$ determined from the method-dependent bandwidths ($9t$) along the high symmetry path in the first Brillouin zone indicated in the inset.}
\label{fig:BandStructure}
\end{figure}

Near graphene, but at low values of the partial pressure of $^4$He gas, interactions between helium atoms can be neglected $(\Vhehe \approx 0)$,  and atoms will be strongly attracted to the centers of the hexagons formed by the graphene lattice as depicted in Fig.~\ref{fig:SingleAtom}(a). The result will be adsorbed atoms with a nearly monolayer density profile \cite{Gordillo:2009jb, Reatto:2013hz,Happacher:2013ht}.  In this limit, the potential $\Vext$ appearing in \Eqref{eq:Ham} can be approximated as being effectively 2D and described by only three numbers: its minimum $(\Vext^{\rm min})$, maximum $(\Vext^{\rm max})$ and saddle-point $(\Vext^{\rm sp})$ values as shown in Fig.~\ref{fig:SingleAtom}(b) with the corrugation fully determined by the smallest two reciprocal lattice vectors of graphene.  These values were recently determined via four different many-body methods \cite{Yu:2021tw}, and the resulting single-particle Hamiltonians were diagonalized to obtain the band structure for a single $^4$He atom on graphene.  The results for the lowest band along a high-symmetry path in reciprocal space are shown in Fig.~\ref{fig:BandStructure}.  

%
%

There is considerable agreement across different methods and the form of the dispersion is  
constrained by symmetries of the adsorption potential.  In the tight-binding approximation on the triangular lattice: 
\begin{equation}
\varepsilon(\vec{k}) - \varepsilon_0 = \! -2t \!  \qty[\!
\cos(k_xa) + 2\cos(\frac{k_xa}{2}) \cos(\frac{\sqrt{3}k_ya}{2})],
\label{eq:tbdispersion}
\end{equation}
where $a=\sqrt{3}a_0$ is the lattice spacing, and $\varepsilon_0$ is an energy offset. Eq.~\eqref{eq:tbdispersion} is plotted as solid lines in Fig.~\ref{fig:BandStructure} and the agreement (combined with a ubiquitous gap in the spectrum \cite{Yu:2021tw}) supports a description of the system in terms of an effective 2D model of $^4$He atoms hopping on the sites of a triangular lattice. Moreover, a hopping (tunneling) matrix element $t \approx \SI{1}{\kelvin}$ can be extracted from the bandwidth (or overlap of maximally localized Wannier functions for helium atoms located at neighboring sites on the triangular lattice \cite{Jaksch:1998,Becker:2010jf}). 

Having understood the single particle problem, it is natural to ask what happens at higher densities (but below first layer completion), where the interaction term $\Vhehe$ in \Eqref{eq:Ham} begins to play a dominant role.  The result will be a low-energy description in terms of one of the simplest and most famous models in condensed matter physics: the Bose-Hubbard model \cite{Gersch:1963ia}.



\section{Correlated Phases of Atoms on Graphene}

The high-energy microscopic Hamiltonian in \Eqref{eq:Ham} describing $^4$He on graphene or graphite has been previously studied, \cite{Corboz2008cb, Bruch:2010fq, Gordillo:2011jb, VranjesMarkic:2013br, Happacher:2013ht}, including identifying the equation of state, and layer completion near filling fraction $f = 2/3$. However, the importance of obtaining an effective  low energy description can be understood as motivating a large family of models to describe the possible properties, quantum phases, and phase transitions of light atoms near 2D materials.  

For $^4$He on graphene, the appropriate description is \cite{Yu:2021tw}:
\begin{equation}
    H  = -t \sum_{\langle i,j\rangle} (b_{i}^\dagger b_j^{\phantom \dagger} + \text{h.c.}) + V  \sum_{\langle i,j\rangle} n_i n_j +  
           V^\prime \!\!\!\sum_{\langle \langle i,j \rangle \rangle} n_i n_j + \dots ,
           \label{eq:BHHamiltonian}
\end{equation}
where $b_i^{\phantom \dagger} (b^{\dagger}_i)$  destroys (creates) an atom at site $i$ of the triangular lattice formed by the graphene hexagon centers occupied by at most $n_i = b_i^\dagger b_i^{\phantom \dagger}$ atom, and  $[b^{\phantom \dagger}_i,b^{\dagger}_j] = \delta_{i,j}$ captures the bosonic nature of $^4$He.  The hopping $t$ is fixed by the lattice potential of graphene (Fig.~\ref{fig:SingleAtom}(a)), and $V, V'$ are the $\langle \text{nearest}\rangle$, and $\langle\langle \text{next-nearest}\rangle\rangle$ neighbor interactions between $^4$He atoms. Interactions beyond second neighbor will be present due to the tails of the dispersion interaction between adsorbed atoms (as indicated by the ellipsis).  

This model is both conceptually and technically different from those usually studied in dilute cold atom systems \cite{Bloch:2008}, where the two-body interaction can be taken to have a zero-range pseudo-potential form, determined by the scattering length.  In that case, the optical lattice wavelength and the atomic confinement scale are much larger than the spatial range of the potential, and consequently the  one- and two-particle length-scales are well-separated during the construction of the associated \emph{soft core} Bose-Hubbard model.  Quantum phase transitions can be observed between superfluid and insulating Mott phases \cite{Bloch:2008, Bakr:2010yv} while more complicated models have been constructed to describe dipolar atomic gases (involving longer-range interactions) and Bose--Fermi mixtures \cite{Lewenstein, Dutta_2015}. In all these cases, as opposed to the He-graphene system studied here, the dominant interaction term is of the soft-core type: $Un_i (n_i-1)/2$ with $U/V$ large but finite, and experimental tunability comes form manipulating the strength of the one-body lattice potential ($t$).  



\begin{figure}[t]
\begin{center}
    \includegraphics[width=\columnwidth]{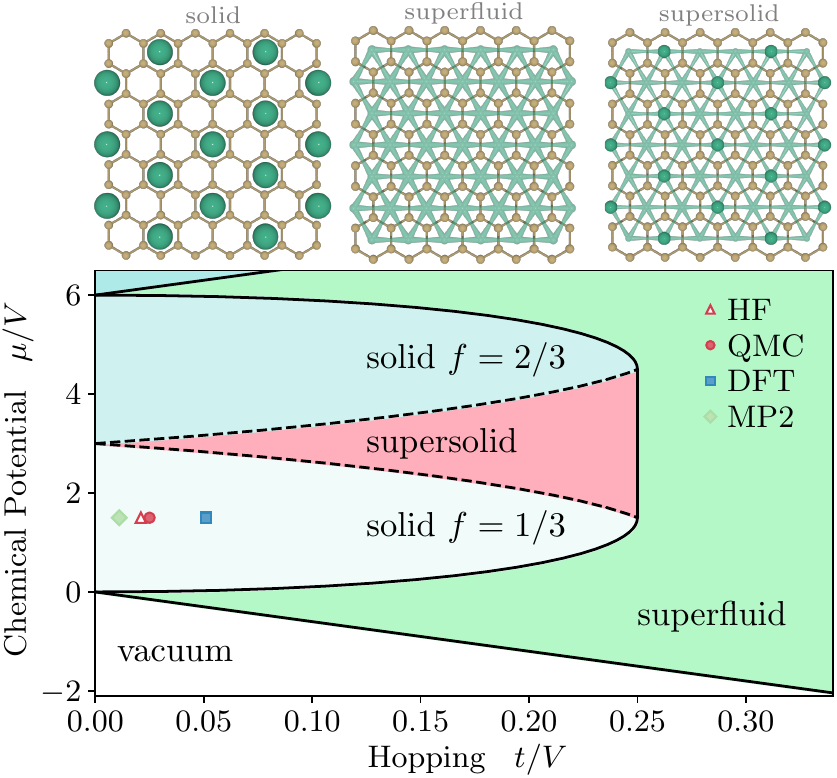}
\end{center}
\caption{Mean field phase diagram for hardcore bosons on the triangular lattice as a function of the chemical potential $\mu$ and hopping strength $t$ measured in units of the nearest-neighbor repulsion $V$.  First-order phase transitions between commensurate solid phases at filling fractions $f=1/3$ (schematic upper left) and $f=2/3$ and a superfluid phase are shown with solid lines.  Continuous second-order phase transitions to a superfluid and supersolid phase (co-existing superfluid and positional order) are indicated with dashed lines with cartoon depictions displayed at the top.  The data points are recent results taken from Ref.~\cite{Yu:2021tw} for four methods: Hartree--Fock (HF), quantum Monte Carlo (QMC), Density Functional Theory (DFT) and M\o{}ller--Plesset (MP2).  They localize the ground state of $^4$He on graphene in the $f=1/3$ commensurate solid phase. All data, code, and scripts needed to reproduce the results are included Ref.~\cite{repo}.}
\label{fig:PhaseDiagram}
\end{figure}

The situation in Eq.~\eqref{eq:BHHamiltonian} is markedly different,  
due to the fact that the adsorbed helium atoms are in a solid-state lattice environment, where the characteristic extent of their atomic wave-functions and the range of the two-body He--He potential are comparable to each other, both on the scale of several angstrom.  Consequently, the one- and two-particle properties cannot be clearly separated,  and the effective model parameters have to be determined in a self-consistent manner via many-body techniques. The model then describes hard-core bosons $(U \approx \infty)$ since the two-body repulsion is extremely strong on the scale of a single graphene adsorption site. The phase diagram of \Eqref{eq:BHHamiltonian}, with only nearest neighbor interactions ($t-V$ model) on the triangular lattice can be analyzed within the mean-field theory \cite{phase-diag,Murthy-phase-diag}, as shown in Fig.~\ref{fig:PhaseDiagram}.  This result is known to be in qualitative agreement with lattice quantum Monte Carlo for hard-core bosons with extended interactions \cite{Wessel:2005ik, Gan:2007zd, Zhang:2011iz}.
%
%
An analysis of the first layer of $^4$He on graphene in the context of the hard-core Bose-Hubbard model \cite{Yu:2021tw} firmly places it in the insulating phase at $1/3$ filling with atoms occupying 1/3 of triangular lattice sites, separated by $\sqrt{3}a_0$, where $a_0 \simeq \SI{1.4}{\angstrom}$ is the carbon--carbon distance.  

Below we briefly summarize how these results were obtained as it highlights the extreme sensitivity of model parameters to physics at the lattice scale.

\begin{figure}[t]
\begin{center}
    \includegraphics[width=\columnwidth]{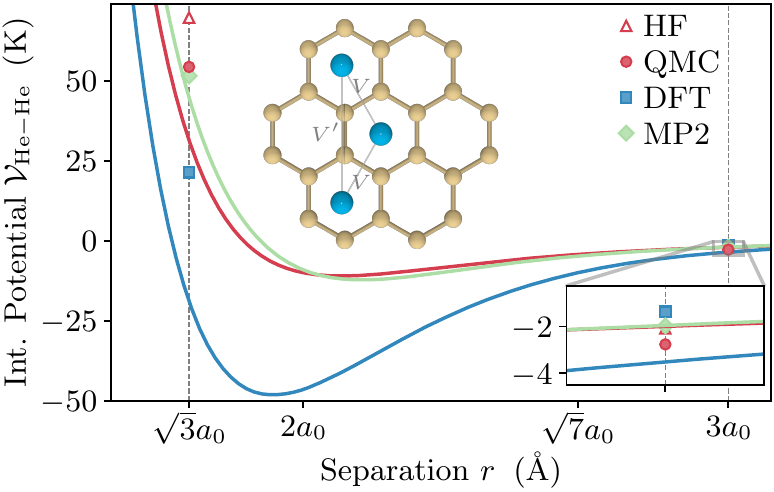}
\end{center}
\caption{The interaction potential energy between two helium atoms in vacuum $\Vhehe$ from the empirical potential introduced in Refs.~\cite{Przybytek:2010ol,Cencek:2012iz} (red line),  density functional theory (DFT, blue line) and M\o{}ller--Plesset perturbation theory (MP2, green line).  
The data points indicate strong corrections with respect to the bare 2-body potential for two nearest neighbor adsorbed atoms separated by $\sqrt{3}a_0$ ($V$), and for the next-nearest neighbor $V^\prime$ at $3a_0$, computed with the many-body methods described in the text.  The upper left inset details the adsorbed configuration, while the lower one shows a zoomed in version of $V^\prime$.}
\label{fig:VHeGraphene}
\end{figure}

A conventional approach (for ultra-cold atoms in optical lattices \cite{Jaksch:1998,Bloch:2008}), to deriving the on-site $(U)$, nearest $(V)$, and next-nearest $(V^\prime)$ interaction terms in effective Bose--Hubbard models, involves convolving the localized single-particle densities on proximate lattice sites computed from single particle Wannier functions (Fig.~\ref{fig:BandStructure}) with the two-body interaction potential.  However, this approach fails here, and does not give physically meaningful results due to the spatial extent of the hard-core of $\Vhehe$ which is on the order of the nearest-neighbor lattice spacing $\sqrt{3}a_0$, (as opposed to being a $\delta$-function) as shown in Fig.~\ref{fig:VHeGraphene}.  
%
%
This immediately leads to $U=\infty$ (two $^4$He atoms cannot simultaneously occupy a single lattice site and still take advantage of the adsorption potential) and the correct effective description is in terms of the \emph{hard core} Bose-Hubbard model defined in \Eqref{eq:BHHamiltonian}. 
To accurately determine the values of the nearest ($V$) and next-nearest ($V^\prime$) interaction parameters, properly computed many-body wavefunctions must be employed to determine the total energy from interactions between atoms at separations $\sqrt{3}a_0$ and $3a_0$ as shown in Fig.~\ref{fig:VHeGraphene}. The renormalization from the bare interaction for point-like helium atoms fixed at these separations can be quantified by the distance between symbols and solid lines.  The resulting model parameters: $V \approx \SI{50}{\kelvin}$ and $V^\prime \approx \SI{-2}{\kelvin}$
are remarkably consistent across different many-body methods with specific values reported in Ref.~\cite{Yu:2021tw}, shown as symbols in Fig.~\ref{fig:PhaseDiagram}, and included in Table~\ref{tab:BHresults}.
\begin{table}
    \renewcommand{\arraystretch}{1.5}
    \setlength\tabcolsep{13pt}
    \begin{tabular}{@{}lllll@{}} 
        \toprule
        Method & $t \, (\si{\kelvin})$ & $V \,(\si{\kelvin})$ & $V^\prime
        \, (\si{\kelvin})$ & ${t}/{V}$ \\ 
        \midrule
        HF & 1.45 & 69.7 & -2.08 & 0.021 \\
        QMC & 1.38 & 54.3(1) & -2.76(2) & 0.025 \\
        DFT  & 1.10 & 21.4 & -1.36 & 0.051 \\
        MP2  & 0.59 & 51.5 & -1.97 & 0.011 \\
        \bottomrule
    \end{tabular}
    \caption{\label{tab:BHresults} Parameters of the effective hard-core Bose-Hubbard model describing $^4$He on pristine graphene defined in \Eqref{eq:BHHamiltonian}.
    We report values from Ref.~\cite{Yu:2021tw} for four different many-body methods: Hartree--Fock (HF), quantum Monte Carlo (QMC), Density Functional Theory (DFT), and M\o{}ller--Plesset perturbation theory (MP2). The hopping $t$ is determined from the single-particle band structure reported in Fig.~\ref{fig:BandStructure}.}
\end{table}
%

\begin{figure}[t]
\begin{center}
    \includegraphics[width=\columnwidth]{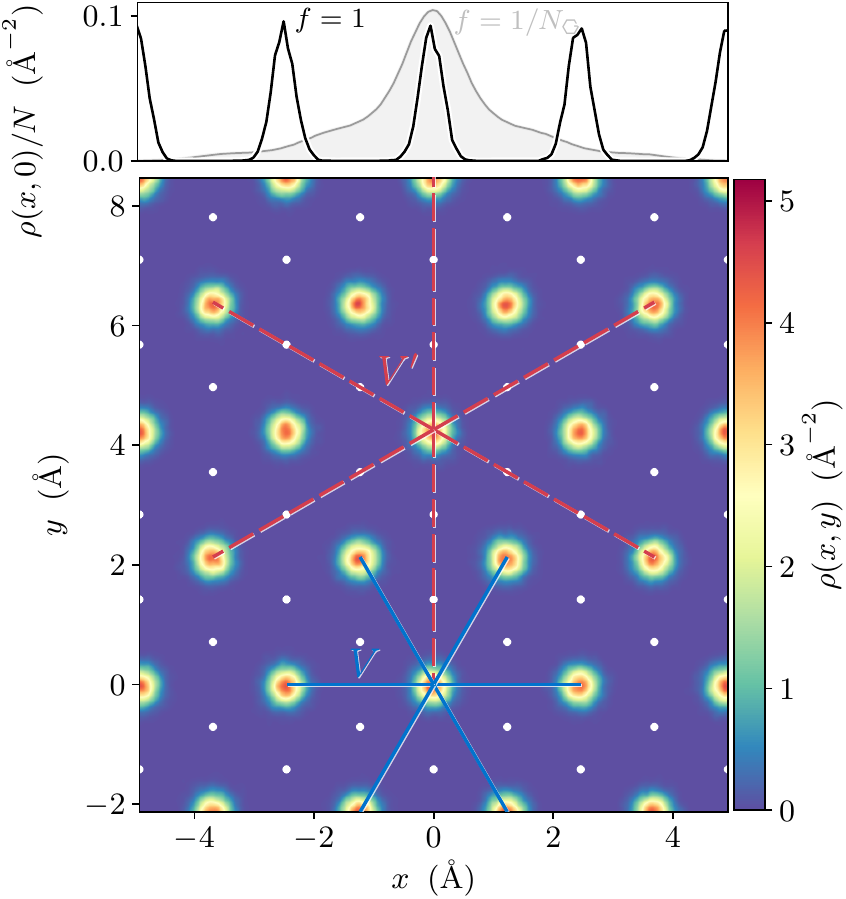}
\end{center}
\caption{The density of $^4$He atoms in the plane $\rho(x,y)$ obtained from
ground state quantum Monte Carlo simulations of $N=48$ $^4$He atoms on $N_\graphene = 48$ adsorption sites.  Only a portion of the cell is shown. The upper plot is a cut through $y=0$ showing the extreme spatial localization of wavefunctions at unit filling in comparison to the extended single-particle ones (grey shaded region) computed by QMC.
Nearest neighbor ($V$) and next-nearest neighbor ($V^\prime$) couplings in the effective Bose--Hubbard description are indicated with solid and dashed lines.}
\label{fig:C1phase}
\end{figure}

The effects of interactions can be clearly seen in the planar density of helium in the first layer $(\rho(x,y))$ shown in Fig.~\ref{fig:C1phase} computed with $T=0$ ground state quantum Monte Carlo \cite{Yu:2021tw}.
%
%
Here, the upper panel compares a cut through the full many-body wavefunction ($\rho(x,0) \propto \Psi(x,0)^2$) with that of a single $^4$He atom on graphene (shaded region), which extends into neighboring lattice sites, consistent with the finite hopping $t$ described above.  This picture illustrates the aforementioned problem of computing $V$ and $V^\prime$ from single-particle Wannier functions that necessitates the use of many-body and ab initio methods.


With the location of the ground state of helium on graphene now pinpointed on the phase diagram of the extended hard-core Bose--Hubbard model (see Fig.~\ref{fig:PhaseDiagram}), we now appeal to previous efforts simulating this much simpler lattice system \cite{Wessel:2005ik, Gan:2007zd, Zhang:2011iz} as opposed to the full 3D microscopic Hamiltonian.  The phase diagram is rich, and includes both superfluid and supersolid phases that are proximate to the filling fraction $1/3$ insulator that may be potentially realizable in this solid state context.  The addition of a next-nearest-neighbor $V^\prime$ further enhances the phase space of interesting physics. We note that this is purely due to the hard-core nature of the interaction $(U=\infty)$ which imposes geometrical and energetic constraints not present in soft-core Bose--Hubbard models with finite $U$ familiar in the context of ultra-cold atoms.

\begin{figure}[t]
\begin{center}
    \includegraphics[width=1\columnwidth]{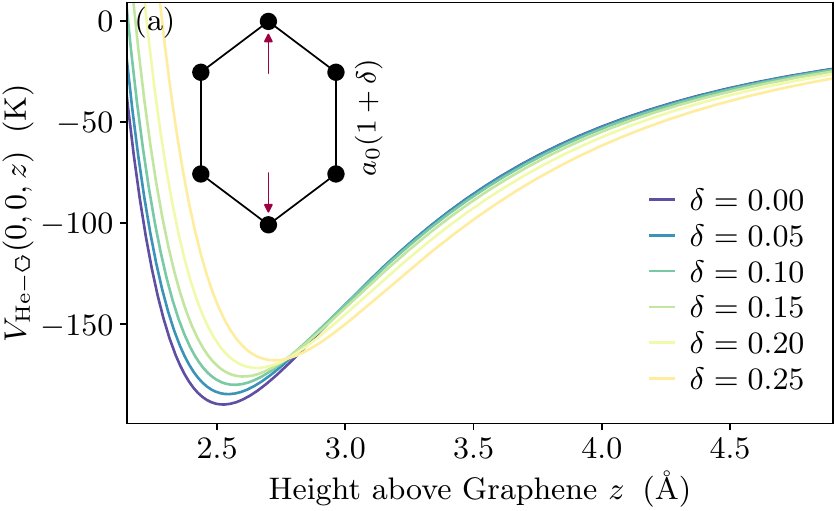}\\[0.1in]
\includegraphics[width=1\columnwidth]{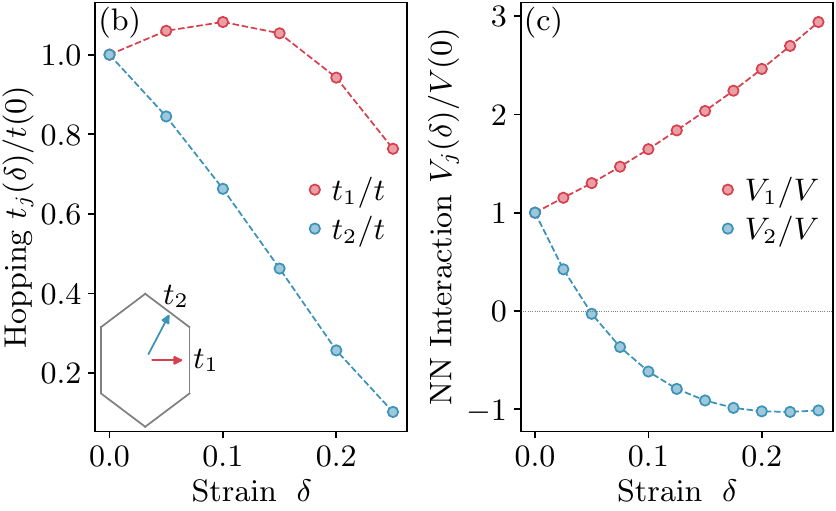}
\end{center}
\caption{Effects of anisotropic strain. (a) The single-particle adsorption potential between $^4$He and graphene for an atom approaching the center of a hexagon as a function of the height $z$ above the sheet.  As strain in the armchair direction of relative strength $\delta$ is applied, the potential minimum is reduced and its position is pushed to larger $z$.  (b) Taking this modified potential into account, the effective hopping $t$ now becomes anisotropic (inset) and the values are strongly affected by strain.  (c) Strain also severely modifies the nearest neighbor adsorption potential $V$, as calculated via density functional theory, and drive a change from repulsive to attractive interactions.}
\label{fig:strain}
\end{figure}

The natural question then arises if $^4$He on graphene can be studied in the laboratory, and whether or not the resulting effective Hamiltonian is tunable.  There exists vast experimental expertise \cite{Dash:1979bk,Crowell:1996kn,McMillan:2005zb} in the preparation and measurement of adsorbed superfluid helium films, and equivalent but non-overlapping efforts to prepare pristine suspended graphene \cite{Meyer:2007ls,Frank:2007ro,Kumar:2018ei}. We propose that the combination of these two research directions, combined with the inherent tunability of 2D systems described in the introduction may provide a novel platform to explore exotic low-dimensional quantum phenomena.  As an tantalizing example of how this tunability could be used to engineer a quantum phase transition, in Fig.~\ref{fig:strain}(a) we show how the single-particle adsorption potential between helium and graphene is effected by the presence of anisotropic strain \cite{Nichols:2016hd} quantified by the length of the carbon-carbon in the armchair direction. 
%
%
The binding energy of a $^4$He atom to the center of a hexagon is reduced and the location of an adsorbed layer is pushed to larger values of $z$ above the graphene.  Such a modified potential has a remarkable effect on the parameters of the effective Bose-Hubbard model (\Eqref{eq:BHHamiltonian}) and in Fig.~\ref{fig:strain} we show that the now anisotropic hopping (b) and nearest neighbor interaction (c) parameters exhibit strong dependence on the amount of applied strain.  While it is not surprising that as the lattice is stretched in the armchair direction, hopping ($t_2$) would be suppressed between a subset of the now further apart anisotropic triangle lattice sites, the behavior of $V_2$ is striking.  At relatively weak strain, density functional theory predicts a crossover from repulsive to attractive interactions at nearest neighbor. This would potentially stabilize a single adsorbed layer of helium at higher filling fraction that could undergo a transition to a purely two-dimensional superfluid, or even anisotropic supersolid phase. Much work remains to be done, and we envision that the search and discovery of such phases will drive new discoveries in low dimensional electronic materials.

We summarize some expectations, and exciting directions, based on general principles described here and known 2D material characteristics,  for possible properties and collective phases  of atoms (in particular He)  on different 2D materials. (1) Based on the analysis above, mechanical strain,  such as uniaxial graphene deformations, can be a powerful tool for studies of zero temperature
quantum phase transitions between insulating and superfluid phases. While insulating phases are expected to be stable at higher temperature, studying superfluidity will require entering a low-temperature regime. (2) Doping of graphene (with electrons or holes)  makes it fully metallic and in general the VDW interaction with helium would
become stronger, due to the larger graphene polarization.  This would would translate into a stronger tendency towards  insulating behavior, but further analysis is needed. (3) The application of a magnetic field  (perpendicular to the membrane) leads to Landau level formation for the graphene electrons and consequently a reduction of electronic polarization \cite{Goerbig2011}.  The resulting weaker adsorption potential will potentially favor superfluidity in a continuously tunable fashion. (4) The application of spatially non-uniform strain, designed specifically to produce a uniform pseudo-magnetic field in a region of space, could also be additionally engineered into a superlattice structure  \cite{Guinea2009}. As even relatively weak strain can produce an enormous effective pseudo-field, this could potentially enhance the behavior suggested in (3). (5) The ability to create and manipulate a Moire superlattice in twisted bilayer graphene leads to the possibility, dependent of  electronic filling, of the graphene quasiparticles switching from insulating to metallic behavior (indeed even  superconducting at low temperatures) \cite{Cao1-2018,Cao2-2018}.
This means that atoms deposited on such a system are certain  to experience dramatically changing  VDW potentials, as a function of the electronic filling in the layers. These represent  only subset of possibilities that would allow for engineering of the microscopic interactions and phases of the $^4$He-graphene system. 

It is also exciting to think that $^3$He can be used instead of $^4$He, which would lead to an effective fermionic model \cite{Todoshchenko:2020}. It is well known that the VDW potentials for $^3$He are quite similar to the bosonic $^4$He, while $^3$He also tends to exhibit rich physics related to its spin and orbital degrees of freedom (such as anisotropic superfluidity, non-trivial topological defects and textures, etc) \cite{volovik2009universe}.

\section{Liquid Films on Graphene and Spinodal De-Wetting}

Next, we extend our discussion to multilayered systems that consist of liquids deposited on top of graphene. Such hybrid arrangements exhibit features 
associated with partial wetting and even  the so-called spinodal de-wetting which reflects a spontaneous surface instability beyond a critical film width.
These effects are due to the 2D nature of graphene and indeed are expected to occur also in other 2D materials as well since they all act as 
``weak adsorbers." Unlike the previous discussion of quantum effects in the first layer which were governed, in large part, by the short-range components
of the  van der Waals (VDW) interactions, the physics of wetting reflects the nature and strength of the  longer-range VDW interaction components between the atoms of the liquid and the atoms and the substrate \cite{Israelachvili,deGennes,Bonn:2009ha}. 2D materials in the suspended configuration are weak adosrbers in a sense that the material--liquid forces may  not 
be strong enough to ensure stable film growth for any film thickness.

\begin{figure}[t]
\begin{center}
\includegraphics[width=0.9\columnwidth]{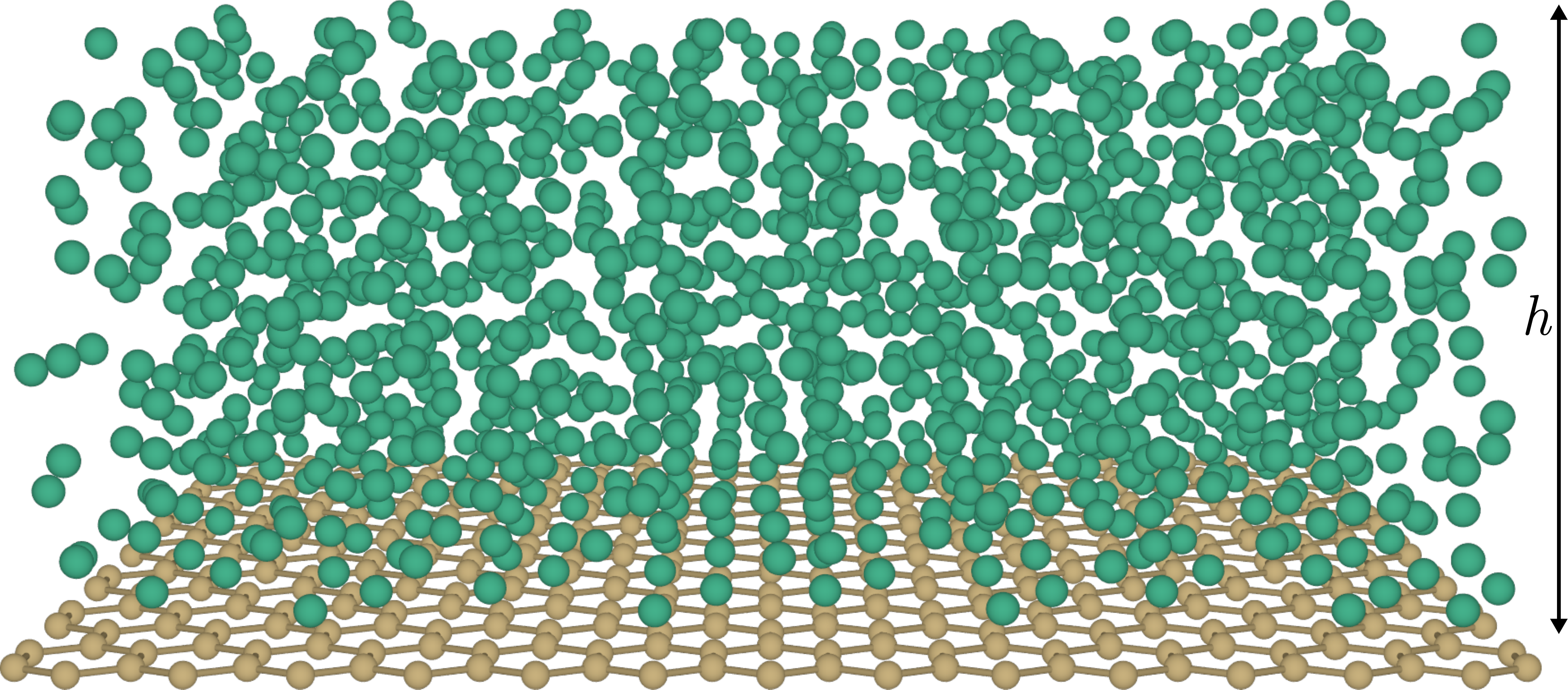}
\end{center}
\caption{Atoms adsorbed on suspended graphene and forming a liquid film of equilibrium width $h$.}
\label{fig:wetting}
\end{figure}

A very successful theory which describes the free energy of  a substrate-liquid-vapor system  for bulk dialectic substrates is the so-called 
 Dzyaloshinskii-Lifshitz- Pitaevskii (DLP) theory \cite{Dzyaloshinskii:2012vn, Dzyaloshinskii:1961vc,LL9,krim,Israelachvili}. It is the standard many-body approach
 to such layered structures 
 and relies on the knowledge of the dielectric functions of all substances involved (the liquid vapor is usually treated as vacuum).
 The discovery of graphene on the other hand  makes it possible to imagine a greater variety of layered arrangements, such as ``suspended" graphene with liquid on top,
 as shown in Fig.~\ref{fig:wetting},  or graphene on a bulk dielectric substrate, with liquid on top of graphene, etc. 
  The conventional DLP theory was recently successfully extended to describe such novel hybrid
 situations  \cite{Sengupta}. Moreover,  the  theory and its predictions were extended to  practically any 2D material, such as the 
 members of the dichalcogenides family (MoSe$_2$, MoS$_2$,  WSe$_2$, WS$_2$) \cite{vdwhetero,DiXiao,Sengupta}. Liquids formed by light atoms, such as
 He,  H$_2$, and N$_2$, can be studied with great theoretical accuracy. Graphene is generally impermeable to light atoms \cite{McEuen,Nair442} while the behavior of
 more complex fluids, like water, is harder to model.

\begin{figure}[t]
\begin{center}
\includegraphics[width=0.95\columnwidth]{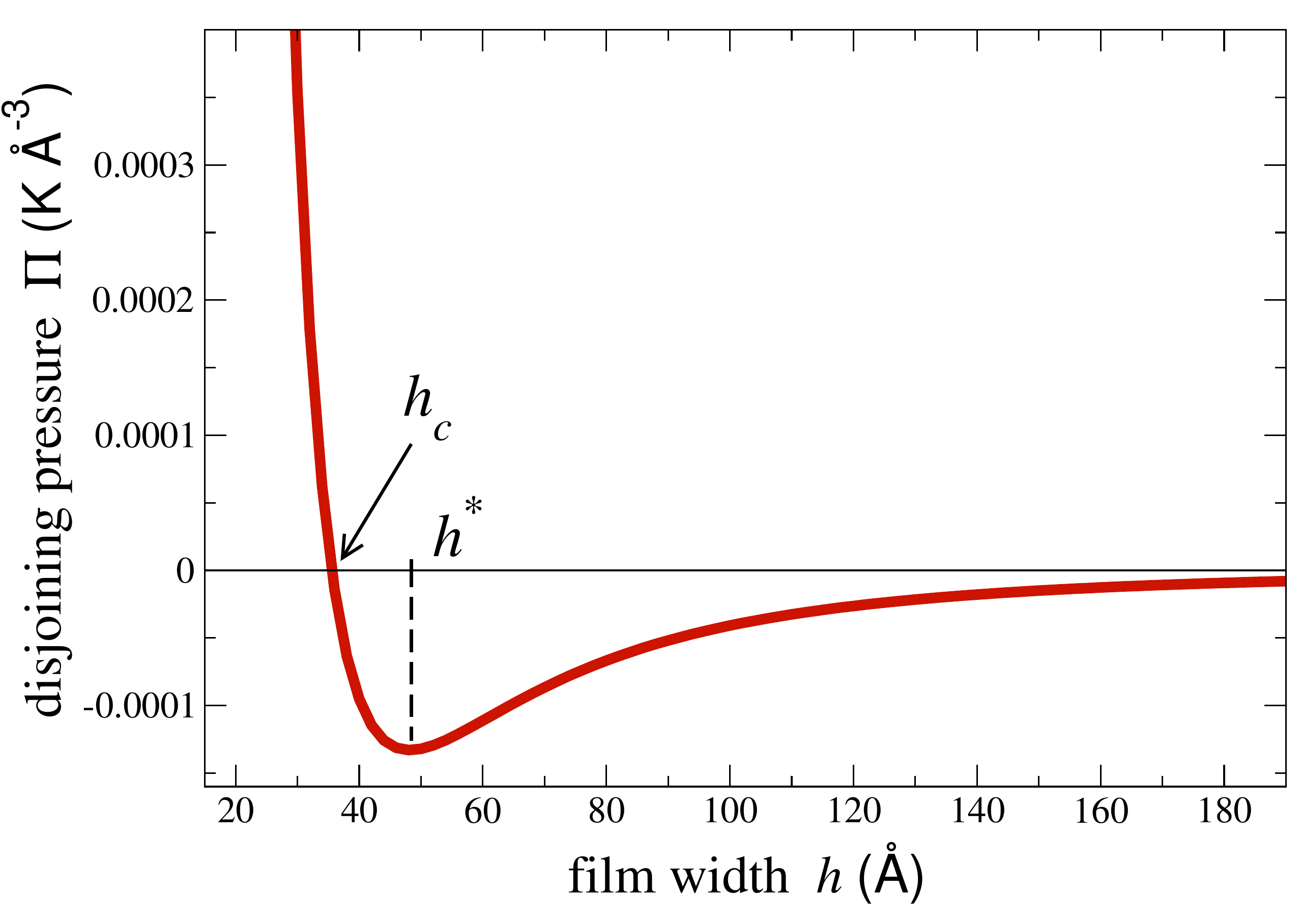}
\end{center}
\caption{Disjoining pressure $\Pi(h)$ for  N$_2$ liquid on top of suspended graphene \cite{Vanegas}.
Light liquids made of He and H$_2$ show similar overall behavior, but with different scales.
The spinodal de-wetting instability takes place for $h>h^*$.
}
\label{fig:dp}
\end{figure}

\begin{figure}[h]
\begin{center}
\includegraphics[width=0.9\columnwidth]{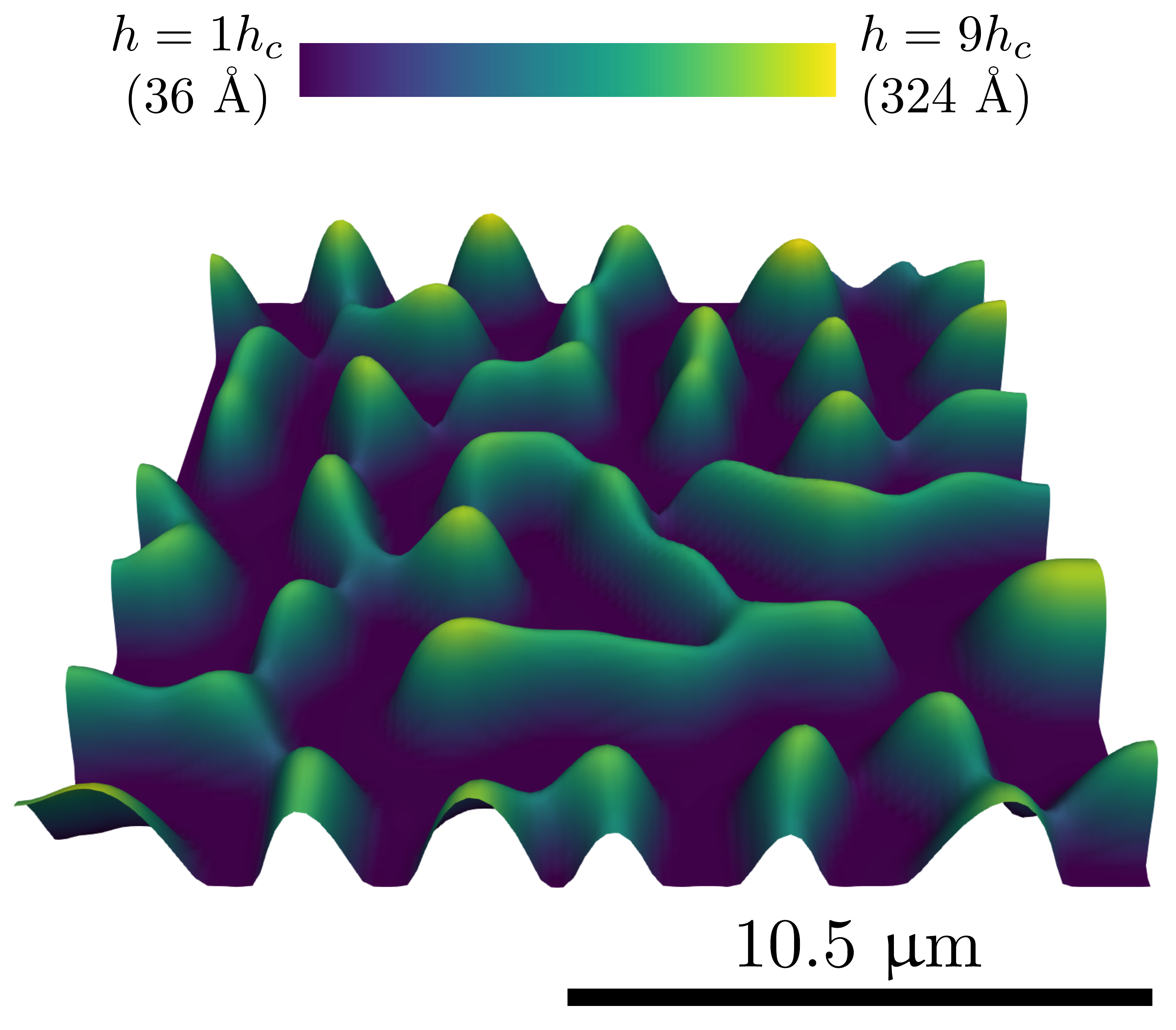}
\end{center}
\caption{Spinodal de-wetting pattern (at intermediate time of the pattern evolution) for N$_2$ liquid on top of suspended graphene, for initial uniform width $h=3h_c > h^*$ \cite{Vanegas}.
``Drier" regions extend
down to $h_c$ and the maximum width fluctuation is $+9h_c$. 
The characteristic spinodal wavelength $\lambda \sim 1 \  \mu\text{m}$.
}
\label{fig:spinodal}
\end{figure}

Let us highlight the two main advantages of using 2D materials for studies of wetting and liquid film growth: (1) As mentioned above, being purely 2D structures, materials like graphene can be engineered and arranged in various configurations. The most exotic of those is the ``suspended" configuration (Fig.~\ref{fig:wetting}).  (2) The polarization function of 2D materials can be calculated with great accuracy. This  in turn leads to an excellent description of VDW forces. Moreover the polarization of graphene reflects its  characteristic Dirac-like electronic dispersion which can be  affected by external factors such as application of mechanical  strain \cite{maria,Nichols:2016hd}, change in the chemical potential (addition of carriers) \cite{Antonio,Kotov}, change in the dielectric environment (i.e.\@ presence of a dielectric substrate affecting screening), etc. The polarization properties of other 2D materials (such as the dichalcogenides family) can also be accurately computed \cite{Sengupta}.  This means that VDW-related properties  and the conditions for film growth can be in principle  effectively manipulated.

Because  graphene is an atomically thin material, it exerts a relatively weak VDW force on the atoms of the film (``weak adsorber"). 
The energy balance governing liquid film growth under equilibrium conditions depends on the balance of VDW interactions (at a given width $h$)
between the atoms of the liquid and the substrate (graphene),  and the atom-atom interactions themselves. In the context of wetting, it is very useful to 
introduce  the concept of disjoining pressure, $\Pi(h)$, which is related to the change (as a function of $h$) of the VDW energy of the system, e.g. in  the configuration shown in Fig.~\ref{fig:wetting}.
The disjoining pressure describes the effective force per unit area between the two boundaries of the system (liquid--vapor and liquid--graphene),
or, equivalently, the difference between the pressures in the finite-width film and the bulk phases \cite{Israelachvili,deGennes,Bonn:2009ha}.
Calculations performed in recent work \cite{Sengupta,Vanegas} show the behavior presented in Fig.~\ref{fig:dp}, which appears to be generic
 for light liquids on graphene.
A change in the sign of  $\Pi(h)$ at a critical $h=h_c$ guarantees a minimum of $\Pi(h)$  at  $h^* > h_c$, with a  
 change in the sign of  $d\Pi(h)/dh$ at $h^*$ (since in the bulk limit $\Pi(h) \rightarrow 0, h\rightarrow \infty$, due to the VDW origin of $\Pi$). 
  While in the range of widths $h_c<h<h^*$ the surface is expected to be metastable, 
 it is known that the regime where  $d\Pi(h)/dh > 0$, i.e. for widths $h>h^*$, corresponds to an  instability \cite{Vrij,Jain,Mitlin,Bonn:2009ha,LL5},
 which in the present context is called spinodal de-wetting. 
The phenomenon of spinodal de-wetting itself has a long history \cite{Vrij, Jain} and has been theoretically
predicted and detected in numerous situations involving polymers, liquid metals, etc. 
\cite{Sharma,Reiter,Vrij,Jain,Herminghaus916,Mitlin,Oron,Xie, Bonn:2009ha, Gentili}. The most important 
feature is the formation of  surface patterns exhibiting characteristic spinodal wavelengths, which  evolve on length and time scales dependent 
on the form of $\Pi(h)$  and 
on the liquid's viscosity and surface tension. The patterns form spontaneously, reflecting the instability of the liquid (width $h>h^*$), 
  subject to any initial surface disturbance.

\begin{table}
    \renewcommand{\arraystretch}{1.2}
    \setlength\tabcolsep{16pt}
    \begin{tabular}{@{}ll@{}} 
        \toprule
        2D Material & $h_c \  (\text{\AA})$ \\ 
        \midrule
        graphene (neutral semimetal)& 300  \\
        doped graphene ($\varepsilon_{F}\!=\!0.3\ \text{eV}$) & 360  \\
         doped graphene ($\varepsilon_{F}\!=\!0.5\  \text{eV}$) & 410  \\
         WS$_2$ (insulator)& 175  \\
          WSe$_2$ (insulator)& 180  \\
           MoS$_2$ (insulator)& 178  \\
            MoSe$_2$ (insulator)& 184  \\
        \bottomrule
    \end{tabular}
    \caption{\label{tab:hc-materials} The length-scale $h_c$ for $^4$He (taken as candidate adsorbate), which corresponds to the change of the  disjoining pressure sign 
    (similarly to  Fig.~\ref{fig:dp} for N$_2$), for various suspended 
    2D material configurations. Here $\varepsilon_F$ is the Fermi energy above the Dirac point in graphene.
     The results are based on calculations in \cite{Sengupta} and the existence of a finite $h_c$ guarantees the development of a spinodalde-wetting instability (at $h^*>h_c$). The values of $h_c, h^*$, and the spinodal wavelengths that follow, are strongly liquid and material dependent. 
    The critical width $h_c$ increases as graphene evolves from neutral to more metallic, while it is  considerably shorter for the members of the dichalcogenides family.  }
\end{table}

Recent work has shown a new path toward creating and controlling spinodal de-wetting patterns for light liquids
forming on suspended graphene and other members of the 2D material family \cite{Sengupta,Vanegas}. 
A typical pattern is shown in Fig.~\ref{fig:spinodal}. The spinodal wavelengths are generally quite long compared to the critical film thickness $h^*$ for spinodal onset (which is up to several hundred $\text{\AA}$),  and range between $\lambda \sim 1 \ \mu \mbox{m}$ and $\lambda \sim 100 \ \mu \mbox{m}$, depending
on the liquid and the material involved. On the solid-state (material) side, there are numerous factors that affect $h_c$ and $\lambda$; for example doping of graphene (adding carriers), application of strain,
replacing graphene with dichalcogenides (which creates a gap in the electronic) spectrum, and various combinations of the above factors.
Table~\ref{tab:hc-materials} presents several examples.
 2D materials are quite special in this regard because any modification of their electronic structure by application of various factors, as outlined above, leads to  well-defined
changes in the polarization properties which in turn affects conditions for film growth within accepted recent theoretical framework \cite{Sengupta,Vanegas}.
Consequently this creates, at least from  theoretical viewpoint, a unique type of functionality and therefore, potentially, 
 a universal theoretical and technological platform for studies of spinodal de-wetting. De-wetting phenomena are known to be of great 
 technological significance in the context of micro-patterning where film structure and length-scale control is very important  \cite{Gentili}.

We also comment on additional factors that could potentially influence the behavior described above. (1) In suspended graphene, acoustic flexural (out-of-plane) phonons
are present which can cause ripples on the scale of tens of \text{\AA} at room temperature \cite{Antonio,Fasolino2007,Katsnelson2012}. Placing graphene on a scaffold, substrate, or under tension,  modifies  this substantially, i.e.\@ quenches the flexural fluctuations. In addition, geometrical effects like bending of the graphene sheet  can cause some modifications of the VDW potential which in principle could be taken into account  \cite{Israelachvili}.
(2) Temperature  can modify  the  VDW potentials, however its effects are generally significant only at larger  distances \cite{Israelachvili,LL9}.
For example the shape in Fig.~\ref{fig:dp} is effectively the same at small or room temperature in the distance range relevant to the problem
(as explained in the supplement to \cite{Sengupta}). The main limiting factor (as far as temperature is concerned) in the analysis is the type of liquid used in calculations: all light liquids mentioned in Fig.~\ref{fig:dp} are stable only well below one hundred degrees Kelvin. It would be interesting to test the 
predictions both  theoretically and experimentally for more complex molecular or multi-component liquids.
Overall we expect the analysis related to the existence of the spinodal de-wetting instability to be quite robust as 
it depends on, and illustrates, the  ``universal" aspects of 2D materials and their interactions with atoms.  Any additional modifications would have to be
done on a case by case basis and reflect the changes  of  material and atomic properties under given experimental conditions.

\section{Concluding Remarks}

We have outlined the physics behind  two important  sets of collective atomic phenomena, where the presence of graphene, and indeed many of the other 2D materials, would make a crucial difference  for  our ability to  ``engineer" an atomic  many-body state with desired characteristics. The field of 2D electronic materials has progressed at an incredible pace since graphene's discovery in 2004, and modern technology allows for extraordinary  level of 
manipulation of  lattice and electronic characteristics. We have argued that atoms could  respond quite readily to changes in these material properties, and thus could arrange themselves in correlated atomic states which previously were not possible to observe and study.  
We hope this theoretical perspective  stimulates further  work  in the field, literally at the interface of atomic and condensed matter many body physics.
%
%
%

\acknowledgments
We dedicate this perspective to our late colleague, Dr. Darren Hitt, former director of the VT Space Grant Consortium.  Darren's leadership, encouragement, and vision to expand the scope of space grant activities in Vermont was essential to forming our interdisciplinary collaboration.

This work was supported, in part, under NASA grant number 80NSSC19M0143.  Computational resources were provided by the NASA High-End Computing (HEC) Program through the NASA Advanced Supercomputing (NAS) Division at Ames Research Center. 

\nocite{apsrev42Control}
\bibliographystyle{apsrev4-2}
\bibliography{refs}

\end{document}